\documentclass[aps, prl, floatfix, twocolumn, showkeys, showpacs]{revtex4} 
\usepackage{amsmath, amssymb, amsfonts}
\usepackage{graphicx}
\usepackage{color}
\usepackage{hyperref}

\newcommand{\figref}[1]{Fig.\,\ref{#1}}

\begin{document}
\title{Constituent mass dependence of transport coefficients in a quark-gluon plasma}

\author{C. E. Coleman-Smith} \email{cec24@phy.duke.edu}
\author{B. M\"uller}
\affiliation{Department of Physics, Duke University, Durham, NC 27708-0305}
\date{\today}

\begin{abstract}

The transverse and longitudinal transport coefficients for jets in a quark-gluon plasma, $\hat{q}$ and $\hat{e}$, are extracted from simulations of the VNI/BMS parton cascade model. We study their variation under the strong coupling and the parton mass in the medium. We show that their ratio is sensitive to the mass of the plasma constituents. Experimental measurements of both quantities could provide new insights into the nature of the quark-gluon plasma as seen by hard probes.

\end{abstract}


\maketitle

%% intro + background
The determination of the transport coefficients of the Quark-Gluon-Plasma (QGP) is a major goal of the LHC and RHIC heavy ion programs. Experimental observations of the suppression of high $p_T$ hadrons at RHIC \cite{PhysRevLett.89.202301, PhysRevLett.88.022301, PhysRevLett.91.072304} and the LHC \cite{Chatrchyan:2011sx, Aad:2010bu} along with the suppression and modification of jets have confirmed the predictions \cite{Gyulassy:1993hr, Thomas1992573, Bjorken:1982tu} of jet-quenching at high momentum. Partons moving through the QGP lose energy and gain momentum perpendicular to their trajectory. 

The details of the mechanisms which cause this suppression remain poorly known, in spite of a great deal of theoretical and experimental effort aimed at elucidating them \cite{Majumder:2010qh, Armesto:2011ht}. The interaction of a hard probe with the QGP medium is traditionally divided into elastic scattering and medium induced radiation. Although this separation may be artificial it is convenient to view the two processes as being independent. The strength of the probe's interaction with the medium is quantified in-terms of the transport coefficients $\hat{q}$ and $\hat{e}$ which represent the average transverse momentum gained and the average energy lost by a hard probe passing through a QGP medium. These can be schematically defined in terms of the differential elastic scattering cross section $d\sigma/dt$,
\begin{align}
  \hat{q} &= \langle \int t \frac{d\sigma}{dt} dt \rangle, \\
  \hat{e} &= \langle \int (E_{f} - E_{i}) \frac{d\sigma}{dt} dt \rangle,
\end{align}
where the angular brackets denote a medium average. More formal definitions can be given in terms of gauge field correlators in QCD \cite{Majumder:2008zg, Majumder:2012sh, Benzke:2012sz}. 

In the limit of infinite probe momentum the quenching process is likely to be well described by radiative proccesses induced by collisions off static color charges. In the opposite limit the dominant energy loss process is likely to be from the recoil of the medium during elastic scatterings. The majority of experimentally studied jet quenching processes lie somewhere between these two limits. Experimental measurements which could distinguish between elastic and radiative modifications to hard probes would not only increase our understanding of the jet quenching process, but also provide an insight into the nature of the QGP medium as seen by hard probes. 

The radiative process and the role of its transport coefficient $\hat{q}$ in jet quenching have been discussed extensively \cite{Baier:2002tc, Arnold:2008vd, Bass:2008rv}. The relative importance of elastic energy loss is less well understood. From kinematic arguments it is clear that the mass of the medium constituents along with the mass of the hard probe will be key in determining how influential energy loss due to elastic processes will be. 

The sensitivity of $\hat{e}$ to the medium mass has been previously discussed by Kolevatov and Wiedemann \cite{Kolevatov:2008bg}. Here we expand upon this work by comparing the scaling of both transport coefficients under variation of the strong coupling constant and the mass of the scattering centers in a relativistic transport model (VNI/BMS) that describes a fully dynamical QCD. We show that the ratio of $\hat{q}$ to $\hat{e}$ is sensitive to the mass of the constituents of the medium. 

%% blurb about model && extraction process?
%%\section{The model}

The VNI/BMS parton cascade model \cite{Geiger:1991nj, Bass:2002fh, ColemanSmith:2011wd, ColemanSmith:2012vr} provides access to the full jet/medium development at a fixed $\hat{q}$ and $\hat{e}$. We run the code in a static uniform-medium mode. Here the medium is modeled as torus of a given radius. Partons that are part of the medium and partons that are part of the jet shower are treated in the same manner, although their origin is tagged within the code. In keeping with the infinite static model of the medium we do not consider hadronization of the jet. We note that hadronization is not necessary for the numerical ``measurement'' of $\hat{q}$ and $\hat{e}$.

The parton cascade model (PCM) is a Monte-Carlo implementation of the relativistic Boltzmann transport of quarks and gluons
\begin{equation}
\label{eqn-boltzmann}
p^{\mu} \frac{\partial}{\partial x^{\mu}} F_k(x, p) = \sum_{i}\mathcal{C}_i F_k(x,p).
\end{equation}
The collision term $\mathcal{C}_i$ includes all possible $2\to2$ interactions and final-state radiation $1 \to n$
\begin{align}
  \label{eqn-pcm-collision}
  \mathcal{C}_i F_k(x,\vec{p}) &= \frac{(2 \pi)^4}{2 S_i} \int \prod_{j} d\Gamma_j | \mathcal{M}_i | ^2 \times \notag\\
  &\delta^4\left(P_{\rm in} - P_{\rm out}\right) D(F_k(x, \vec{p})),
\end{align}
$d\Gamma_j$ is the Lorentz invariant phase space for the process $j$, $D$ is the collision flux factor and $S_i$ is a process dependent normalization factor. A geometric interpretation of the total cross-section is used to select pairs of partons for interaction. Between collisions, the partons propagate along straight line trajectories.  The QGP medium is simulated as a box of thermal quarks and gluons generated at some fixed temperature. Periodic boundary conditions are imposed on the box, whose size is selected to be large enough that if a simulated jet wraps around it will not interact with its own wake.

%% actually use the asymptotic masses instead
The medium partons are initialized with a thermal distribution of energies with masses given by the asymptotic Hard Thermal Loop (HTL) \cite{PhysRevD.45.R1827, Blaizot:2005wr}  result for partons in a thermal plasma with zero chemical potential, 
\begin{equation}
m^2 = k g^2 T^2, \quad k = 
\begin{cases}
\scriptsize{\frac{1}{6}}N_c + \scriptsize{\frac{1}{12}N_{f}} & \text{gluon} \\
\scriptsize{\frac{1}{4}}C_{F}  & \text{quark}
\end{cases}
\end{equation} where $N_{f}$ is the number of flavors present in the plasma,  $N_c$ is the number of colors and  $C_{f} = (N_c^2-1)/2N_c$ is the Casimir invariant for the fundamental representation of SU($N_c$). To investigate the role of the mass of the medium partons in the jet quenching process we introduce an additional scaling constant $\mu_{s}$ such that 
\begin{equation}
m^2 = k \mu_{s}^2 g^2 T^2.
\end{equation}
This scaling is applied equally to the mass of all the quark flavors and the gluons making up the medium.

%% ratio = 4T, why? 

%% extraction process
To study the transport coefficients we run the model without radiative processes. We insert a light quark into a box at $T=350$~MeV and propagate it for a fixed distance of $L=4$~fm with a baseline strong coupling $\alpha_{s} = 0.3$. We extract $\hat{q}$ from VNI/BMS by computing the average amount of transverse momentum broadening of the parton per unit length $L$ over some number of elastic collisions with the medium $N_{\rm coll}$
\begin{equation}
  \label{eqn-qhat-pcm}
  \hat{q} = \frac{1}{L}\sum_{i=1}^{N_{\rm coll}} \Delta p_{T,i}^{2}.
\end{equation}
We extract the elastic scattering transport coefficient as 
\begin{equation}
  \label{eqn-ehat-pcm}
  \hat{e} = \frac{1}{L}\sum_{i=1}^{N_{\rm coll}} \Delta E_{i},
\end{equation}
where $\Delta E_{i}$ is the energy change of the parton during the $i$th collision.

\begin{figure}[htb]
\includegraphics[width=0.35\textwidth]{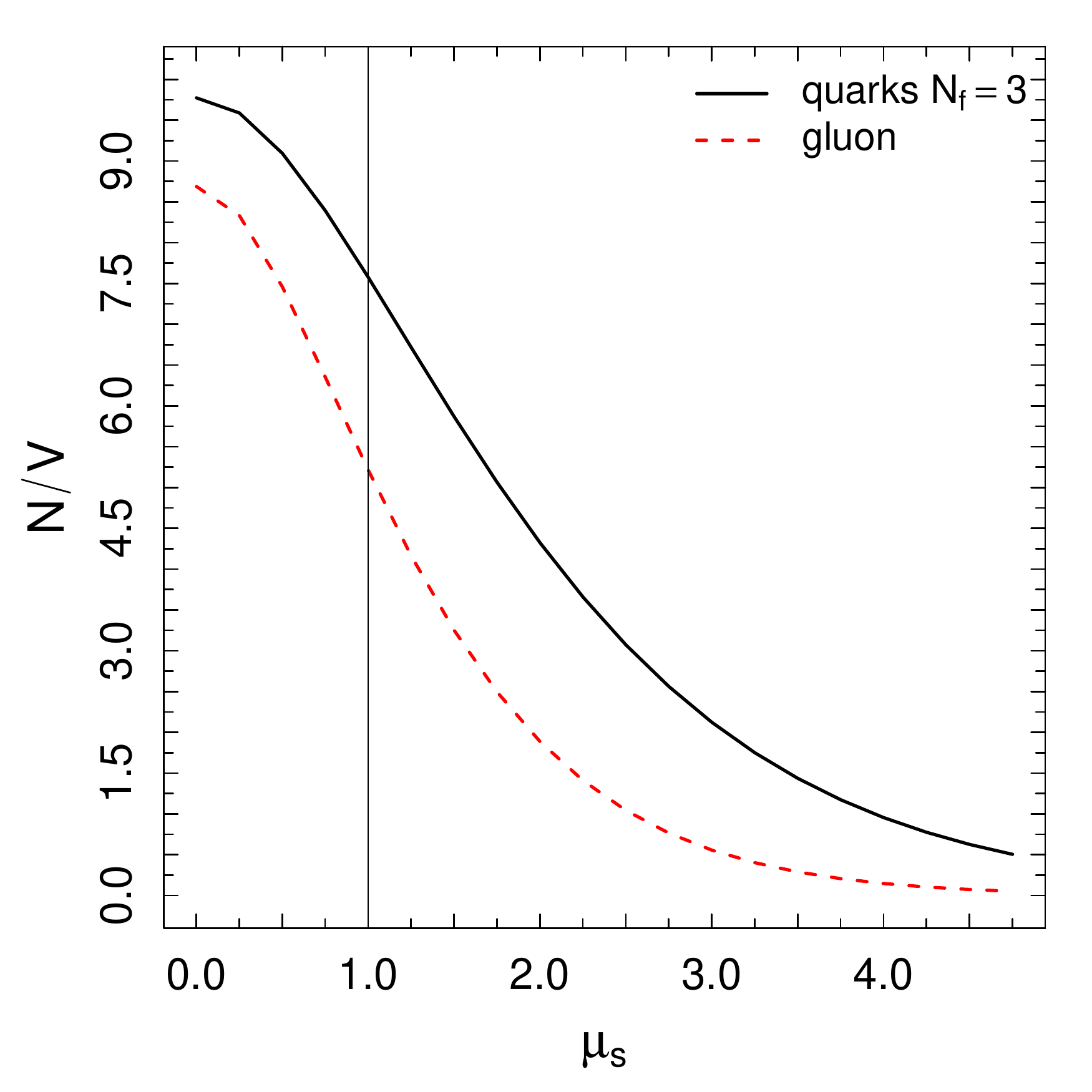}
\caption{(Color Online) The medium number density \eqref{eqn-density} as a function of the medium constituent mass scale $\mu_{s}$, for a medium at $T=350$~MeV with $N_{f} = 3$ and $\alpha_{s} = 0.3$. The vertical line marks the values of the number density for no scaling of the constituent masses ($\mu_{s}=1$).}
\label{fig:density-scale}
\vspace{-5mm}
\end{figure}

The parton density in the box is computed for an ideal gas of ultra-relativistic massive particles, 
\begin{equation}
  \label{eqn-density}
  \frac{N}{V} = \frac{g T m^2 }{2 \pi^2} K_2(\frac{m}{T}),
\end{equation}
where $g = 2(N_{c}^2-1)  = 16$ for gluons, $g = 2 N_{c} N_{f}  = 18$ for quarks and $K_n(x)$ is the modified Bessel function of the second kind \cite{Abramowitz:1972}. The medium density depends strongly upon the mass scale $\mu_{s}$ as is shown in \figref{fig:density-scale}. We will scale out the medium density from $\hat{q}$ and $\hat{e}$ when investigating their dependence upon $\mu_{s}$, and we note that the density drops out of the ratio $\hat{q}/\hat{e}$.

\begin{figure}[htp]
\includegraphics[width=0.35\textwidth]{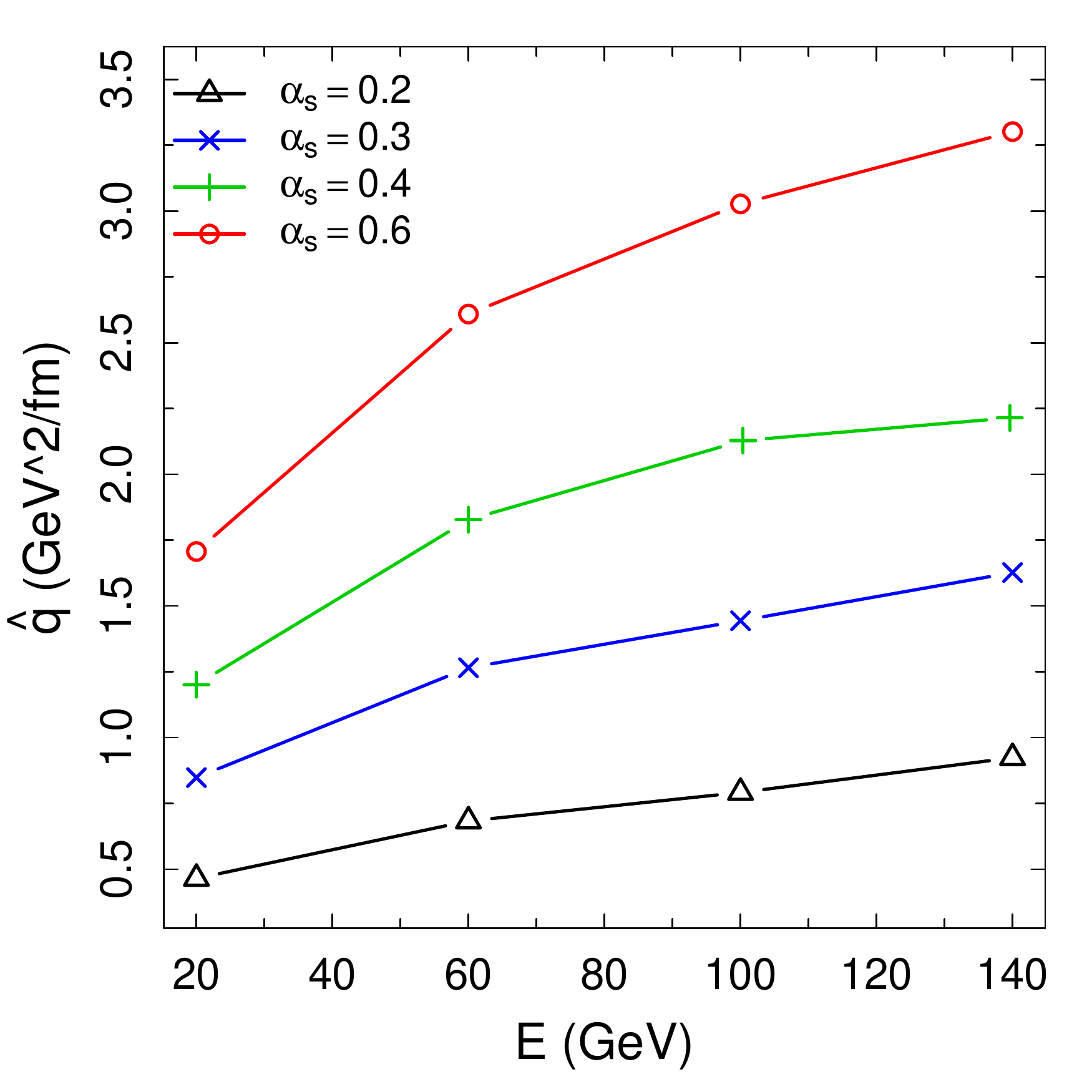}
\includegraphics[width=0.35\textwidth]{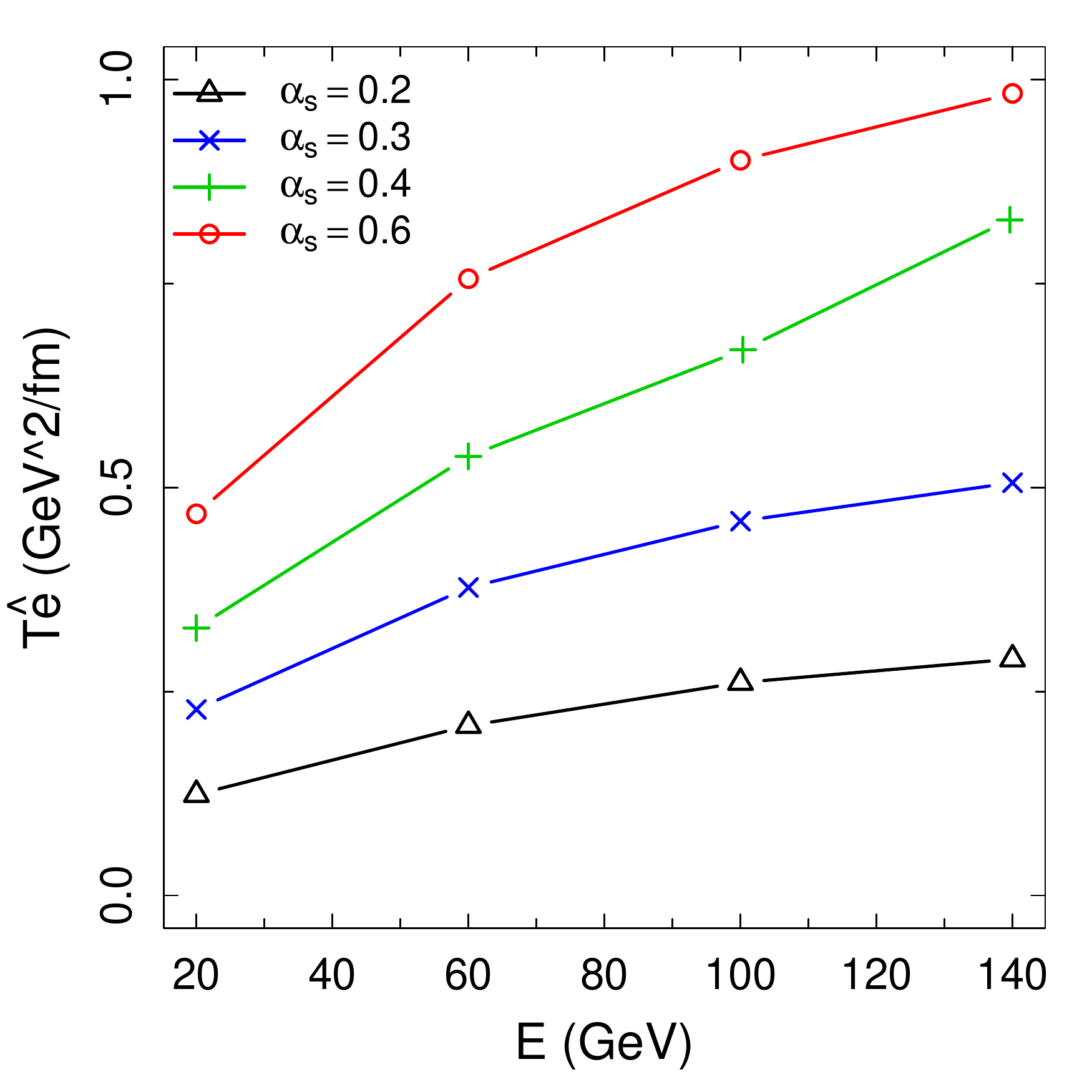}
\caption{(Color Online) The scaling of the transport coefficients with strong coupling $\alpha_{s}$, for a medium at $T=350$~MeV and $\mu_{s} = 1$.}
\label{fig:alpha-scale}
\vspace{-5mm}
\end{figure}

In \figref{fig:alpha-scale} we show the scaling of $\hat{q}$ and $\hat{e}$ with the strong coupling $\alpha_{s}$ as a function of the probe energy. Both transport coefficients scale roughly quadratically with $\alpha_{s}$ as would be expected from the perturbative massless medium results \cite{Arnold:2008vd, Braaten:1991we, Peigne:2008nd} 
\begin{align}
  \label{eqn-qhat}
  \hat{q}(T) &= 4\pi C_R \alpha_s^2 \mathcal{N}(T) \ln\left(\frac{q_{\rm max}^2}{m_D^2} + 1\right), 
  \notag\\
  m_D(T)^2 &= \left(\frac{N_c}{3} + \frac{N_{f}}{6}\right) g^{2}T^{2}, 
  \notag \\
  \mathcal{N}(T) &= \frac{\zeta(3)}{\pi^2}\left( 2N_c+\frac{3}{2}N_f \right)T^3, 
  \notag \\
  q_{\rm max}^2(E,T) &= E T.
\end{align}
and 
\begin{equation}
  \label{eqn-ehat}
  \hat{e}(T) = \frac{4\pi\alpha^2 T^2}{3} \left(1+\frac{N_{f}}{6}\right)\ln\left(\frac{ET}{m_D^2}\right)
\end{equation}

Here we have held the medium mass scale $\mu_s = 1$ and the ratio of $\hat{q}$ to $\hat{e}$ is approximately $4$.

\begin{figure}[hbtp]
\includegraphics[width=0.35\textwidth]{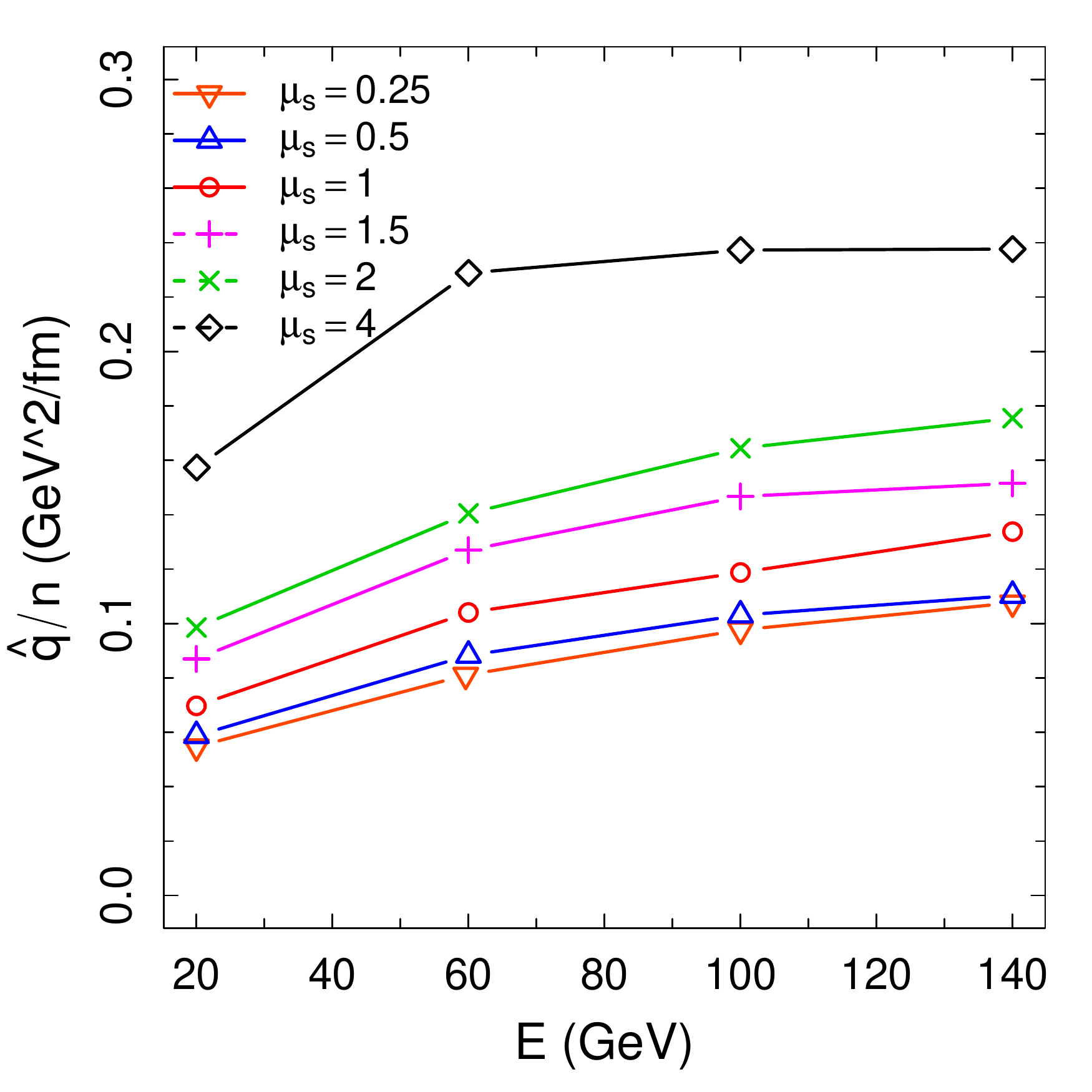}
\includegraphics[width=0.35\textwidth]{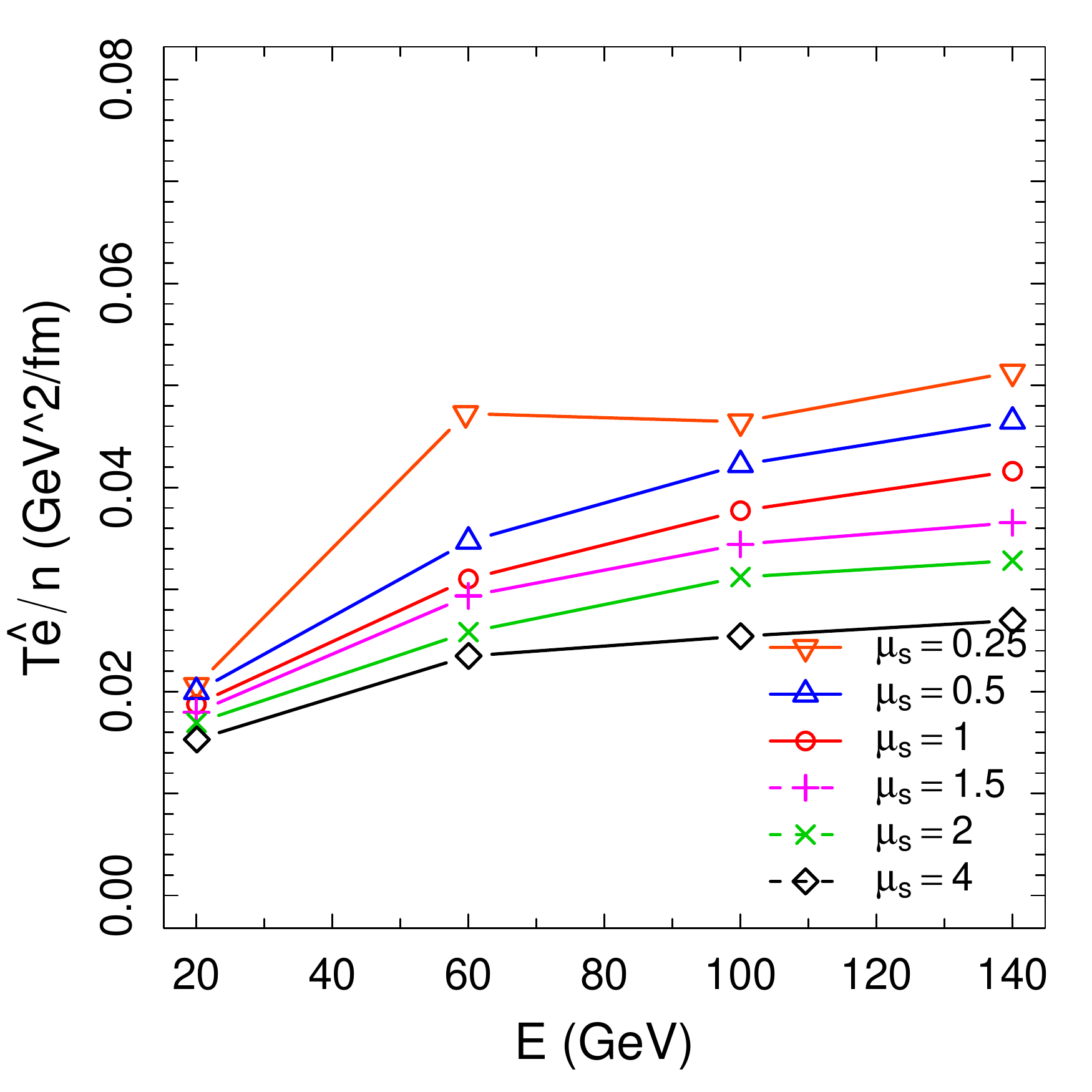}
\caption{(Color Online) The scaling of the transport coefficients with the medium mass scale $\mu_s$, here $\hat{q}$ and $\hat{e}$ have been scaled by $n(\mu_s) = N/V(\mu_s)$ as determined by \eqref{eqn-density}.}
\label{fig:ms-scale}
\end{figure}

In \figref{fig:ms-scale} we hold the strong coupling fixed $\alpha_{s} = 0.3$ and vary the medium mass scale $\mu_{s}$. Note that we have scaled out the dependence upon the medium density in these figures. As we increase the medium mass scale $\hat{q}$ increases. This arises purely from the kinematics of the scattering. Scatterings off heavier medium constituents favor transverse momentum transfer over recoil. The average energy loss decreases with $\mu_{s}$, a light medium will transfer energy from an incoming probe into recoil of the constituents. 

\begin{figure}[htp]
\includegraphics[width=0.35\textwidth]{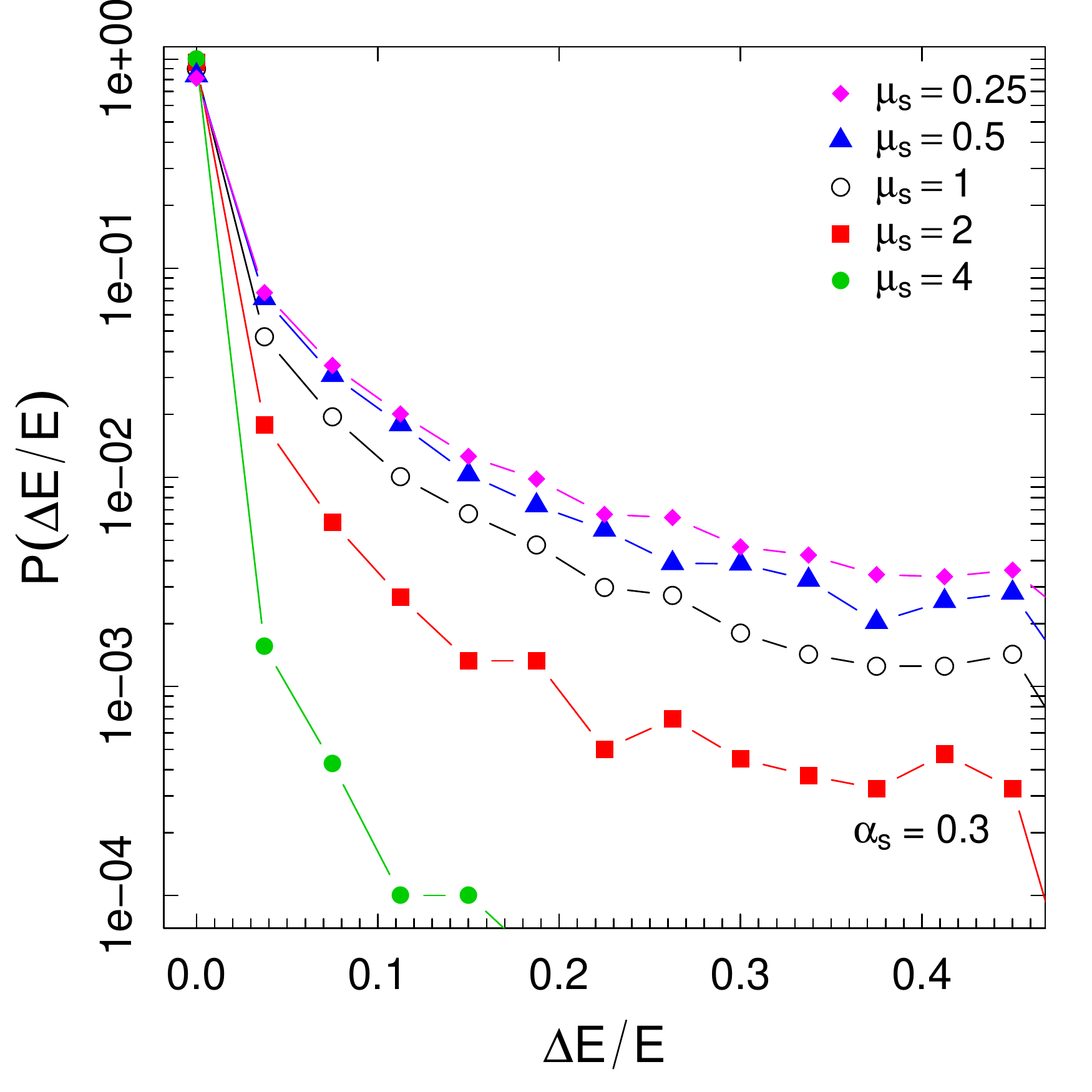}
\caption{(Color Online) The probability of total energy loss for a quark with $E=100$~GeV traveling through $4$~fm of medium at a $T=350$~MeV and $\alpha_{s}=0.3$ as a function of the medium mass scaling parameter $\mu_s$.}
\label{fig:de-distribution}
\end{figure}

\figref{fig:de-distribution} shows the distribution of elastic energy loss as a function of the medium mass scale $\mu_{s}$, here the influence of varying $\mu_{s}$ can be observed. The lighter medium leads to greater recoil and an enhanced tail in the $\Delta E/E$ distribution compared to the baseline, the heavier medium leads to a suppression of the spectrum.

%% ccs, plot P(dpt / pt) also

\begin{figure}[htp]
\includegraphics[width=0.35\textwidth]{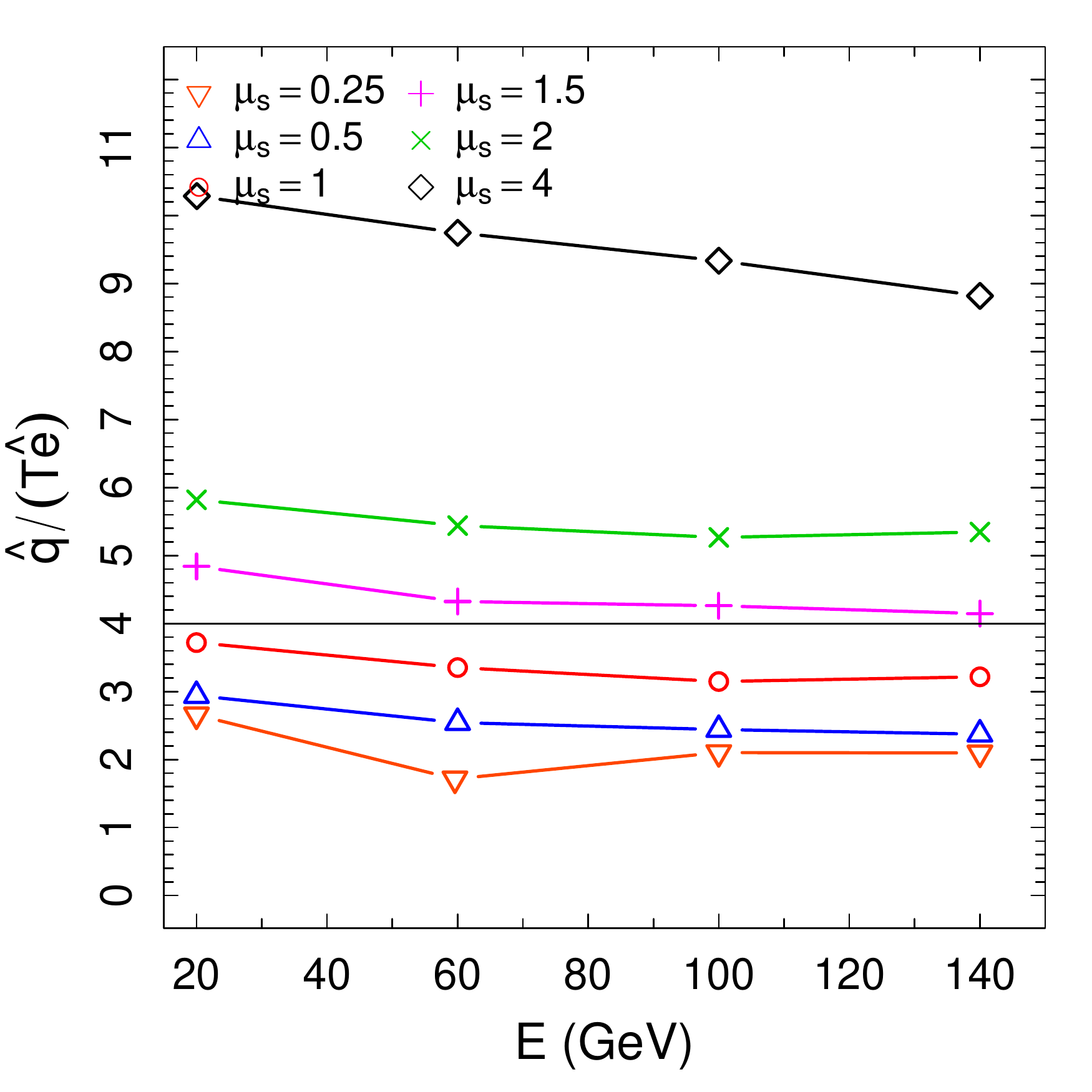}
\caption{(Color Online) The ratio of the transport coefficients as a function the medium mass scale $\mu_{s}$ for $T=350$~MeV and $\alpha_{s}=0.3$.}
\label{fig:ratio}
\vspace{-3mm}
\end{figure}

\figref{fig:ratio} shows the ratio of the transport coefficients as a function of the probe energy. The ratio scales roughly linearly with the medium mass $\mu_{s}$. Heavy constituents promote transverse momentum transfer and so radiative processes, lighter constituents favor energy transfer in elastic scattering. The dependence upon the medium density cancels in the ratio of the transport coefficients, thus the ratio is only sensitive to the mass of the medium constituents. Experimental measurements of this ratio would provide a direct measure of the mass of the medium constituents as seen by hard probes. These measurements could be carried out over a range of probe virtualities, by comparing heavy and light quark jets at the LHC with those at RHIC. An experimental determination of the ratio $\hat{q}/\hat{e}$ would give vital insight into the quasi-particulate nature of the QGP.

%% conclusions

In summary, we have confirmed by an explicit numerical simulation that the ratio of the jet transport coefficients $\hat{e}$ and $\hat{q}$ is sensitive to the mass of the medium constituents. Our result raises the possibility of determining the masses of quasiparticles as encountered by a hard probe from experimental data. The validity of our calculation is limited for high quasiparticle masses as we use the same screened cross-sections for all mass scales. Also, our model does not include quasiparticle breakup or excitation, which are likely to contribute to inelastic energy loss in the high mass region \cite{Shuryak:2004ax}.

Experimental results could be obtained by constraining the values of $\hat{q}$ and $\hat{e}$ through the systematic comparison of jet-quenching codes to multiple sets of experimental observables across a range of scales. E.~g., jet broadening, determined through back-to-back (gamma-) hadron correlations \cite{Wang:1996yh}, could give direct measurements of  $\hat{q}$. Measurements of jet energy as a function of cone angle for identified photon jets may provide measurements of the amount of energy lost from the jet to the medium, allowing for an estimate of $\hat{e}$.  These measurements would be an important step on the road towards precision measurements of the transport properties of hard probes in the QGP, just as measuring the shear viscosity to entropy ratio \cite{Song:2010mg} has provided a focus for the precision measurement of bulk properties.

\begin{acknowledgments}
We acknowledge support by DOE grants DE-FG02-05ER41367 and DE-SC0005396. This research was done using computing resources provided by OSG EngageVo funded by NSF award 075335. 
\end{acknowledgments}

\bibliographystyle{apsrev4-1}
\bibliography{./qhat-ehat}{}

\end{document}